\documentclass[12pt,preprint]{aastex}
\usepackage{bbm}
\usepackage{mathrsfs}

\newcommand\beq{\begin{equation}}
\newcommand\eeq{\end{equation}}
\newcommand\bea{\begin{eqnarray}}
\newcommand\eea{\end{eqnarray}}
\newcommand\bal{\begin{align}}
\newcommand\eal{\end{align}}

\newcommand\h{\mathbbm{h}}

\newcommand\bk{\bold{k}}

\renewcommand\bal{\mbox{\boldmath$\alpha$}}

\begin{document}

\title{A natural origin of primordial density perturbations}

\author{Richard Lieu\altaffilmark{1} and T.W.B. Kibble\altaffilmark{2}}

\altaffiltext{1}{Department of Physics, University of Alabama,
Huntsville, AL 35899.}
\altaffiltext{2}{Blackett Laboratory,
Imperial College, London SW7~2AZ, U.K.}

\begin{abstract}

We suggest here a mechanism for the seeding of the primordial
density fluctuations.  We point out that a process like reheating at
the end of inflation will inevitably generate perturbations, even on
superhorizon scales, by the local diffusion of energy. Provided that
the reheating temperature is of order the GUT scale, the density
contrast $\delta_R$ for spheres of radius $R$ will be of order
$10^{-5}$ at horizon entry, consistent with the values measured by
\texttt{WMAP}.  If this were a purely classical process,
$\delta_R^2$ would fall as $1/R^4$ beyond the horizon, and the
resulting primordial density power spectrum would be $P(k) \propto
k^n$ with $n=4$.   However, as shown by Gabrielli et al, a quantum
diffusion process can generate a power spectrum with any index in
the range $0<n\leq 4$, including values close to the observed $n=1$
($\delta_R^2$ will then be $\propto 1/R^{3+n}$ for $n<1$ and $1/R^4$
for $n>1$).  Thus, the two characteristic parameters that determine
the appearance of present day structures could be natural
consequences of this mechanism.  These are in any case the minimum
density variations that must have formed if the universe was rapidly
heated to GUT temperatures by the decay of a `false vacuum'.  There
is then no \emph{a priori} necessity to postulate additional (and
fine tuned) quantum fluctuations in the `false vacuum', nor a
pre-inflationary period.  Given also the very stringent
pre-conditions required to trigger a satisfactory period of
inflation, altogether it seems at least as natural to assume that
the universe began in a flat and homogeneously expanding phase.

\end{abstract}

\

The large scale homogeneity and flatness of the early Universe as
revealed by \texttt{COBE} and \texttt{WMAP} observations (Smoot et
al 1992, Bennett et al 2003, Spergel et al 2007, Hinshaw et al 2009)
of the cosmic background radiation (CBR),  together with the
formation of the observed large-scale structures, present three of
the greatest challenges to contemporary cosmology. The hypothesis of
inflation (Guth 1981, Albrecht \& Steinhardt 1982) offers an ingenious
solution to these problems.

In this Letter we wish to focus our attention upon the supremely
important problem of structures.  An observational result of
\texttt{WMAP} that may be interpreted as supportive of conventional
models of inflation is the Harrison-Zel'dovich (HZ) power spectrum
of primordial density modes in Fourier space,
 \beq P(k) = |\delta_k|^2 \propto k^n, \eeq
with $n \approx$ 1.  Provided $P(k)$ `levels off' at $k\approx 1/r_H$
where $r_H$ is the comoving radius of the horizon, the density contrast
$\delta_R$ within a sphere of radius $R$ is given by (see e.g.\ Gabrielli
et al 2002)
 \beq \delta_R^2 = \frac{\langle\delta u^2\rangle}{\bar u^2} =
 \frac{1}{V^2}\int d^3 \bk |W(k)|^2 P(k) \propto
 \frac{1}{R^{3+n}},~n<1;~{\rm and}~\frac{1}{R^4},~n>1, \eeq
where $\bar u$ is the mean density of the universe, $\bar u +\delta u$ the
mean density within the sphere, $V=4\pi R^3/3$, and $W(k)$ is the three
dimensional Fourier transform of the top-hat window function $W(r)$,
defined as $W(r)=1$ when $ r\leq R$ and $W(r)=0$ when $r>R$.

The power spectral index of $n\approx1$, at least from the viewpoint
of the ensuing $\delta_R^2 \propto 1/R^4$ scaling of real space
density variations, is clearly indicative of a higher degree of
spatial homogeneity than a Poisson distribution where $n=0$ and
$\delta_R^2 \propto 1/R^3$.  What is the origin of such an index?
Zel'dovich (1965), and Zel'dovich and Novikov (1983) were among the
first to point out that small and causal displacements of particles
from their original positions in a super-ordered state can only lead
to an index of $n\geq 4$.  Cosmologists then had to resort to more
exotic explanations of $n\approx 1$, e.g.\ perturbations generated
by quantum fluctuations during inflation.  More recently, however,
Gabrielli et al (2004) noted that the conclusion of Zel'dovich et
al, while valid classically, is too restrictive in general: causally
limited quantum interactions have access to the broader range,
$n>0$.  This then opens the possibility of a novel and very natural
kind of explanation for the observed spectral index $n\approx1$.

Let us first be clear about our choice of gauge here and after.
Since superhorizon densities are ambiguously defined, eq.~(2) is
valid only in the synchronous gauge, where the amplitude evolution
of the seed w.r.t.\ $t$ occurs at the rate $\delta_R \propto
r_H^2(t)$ so long as $R> r_H$ and $\delta_R \ll 1$ (section 9.3.6,
Kolb \& Turner 1990). We are interested here in \emph{superhorizon}
evolution, i.e., for a given scale $R$, the period during which $r_H
< R$, between the epoch $t=t_{{\rm exit}}$ when it exited the
horizon during the inflationary era, and the epoch $t=t_{{\rm
entry}}$ at which it re-entered during the radiation (or matter)
dominated era.  As is well known, if the linear evolution equations
are valid throughout this period, then the amplitude $\delta_R$ will
have the same value at $t_{{\rm entry}}$ as it did at $t_{{\rm
exit}}$.  At both limits $\delta_R$ has physical significance and is
unambiguously defined, although in the intervening period it is
gauge-dependent.  Moreover, this `horizon-crossing' value of
$\delta_R$ also does not vary significantly with $R$.  It is
determined observationally by the low-harmonic temperature
anisotropy of \texttt{WMAP} as $\delta_R (R=r_H) \approx$ 3 $\times$
10$^{-5}$ (e.g.\ Bennett et al 2003), although the final four year
\texttt{COBE} data favored the slightly lower number of 2 $\times$
10$^{-5}$ (Peacock 1999).  Beyond the horizon, the observations tell
us that the scaling of $\delta_R$ is close to
 \beq \delta_R \approx 10^{-5}
 \left(\frac{r_H}{R}\right)^2~{\rm for~all~}R \geq r_H. \eeq
The constancy of $\delta_R (t_{{\rm entry}})$ is conventionally taken
to be a reflection of the very slowly changing conditions during
inflation.

In spite of the aforementioned milestone achievements, there remain
significant uncertainties and weakly justified tenets in the
inflation model, most of which have to do with the fine tuning of
the initial conditions (see Padmanabhan 1993 for an account of many
of them). Thus e.g.\ for every scale to exit with nearly the same
amplitude, the Hubble parameter and the time derivative $\dot\phi$
of the scalar field $\phi$ that drove inflation must be very
precisely constant during the horizon exits of all density-contrast
modes relevant to structure formation; indeed modifications of the
basic scenario which lead to different patterns of spatial density
fluctuations do exist (e.g. Kawasaki et al 2003, Hall et al 2004,
Yamaguchi \& Yokoyama 2004).  Secondly, to trigger inflation the
pre-inflationary universe must already be homogeneous on
superhorizon scales (Vachaspati \& Trodden 2000).  Thirdly, the
formula for these delicately constructed zero-point fluctuations has
both infra-red and ultra-violet divergences, and regularization is
needed on each end.  Fourthly, the result depends on a choice of
initial state, usually assumed to be some kind of `vacuum state',
but this is somewhat arbitrary since there is no unambiguous
definition of the vacuum in a curved background.  Lastly, most
scalar field theories of GUT phase transitions involve too strong a
coupling constant; as a result, they yield an exit amplitude larger
than $10^{-5}$ by orders of magnitude.  Hence, the claim that
inflation `solved' the problem of structure formation by
`predicting' the observed density seeds may have been overstated.

The purpose of this paper is to propose a way of improving the
situation. The point to be demonstrated is that the origin of
structures does not have to be the variance in the vacuum
expectation value of $\phi$ at all: even for a completely classical,
uniform $\phi$ the generation of a primordial density pattern with
the observed amplitude is more than a mere possibility; it is
perhaps even an inevitability.  Central to our argument is a revisit
of the $\delta_R(t_{{\rm entry}}) = \delta_R (t_{{\rm exit}})$
relation, which is normally seen as the reason why a classical
vacuum cannot seed structures.  That is, it is often argued that
causal processes cannot create or modify density contrasts on
superhorizon scales. Yet this is only approximately true. Particles
generated during the reheating process can propagate some distance
and so either penetrate or leave a chosen superhorizon-size sphere,
though of course this can only happen for particles lying within a
thin shell. Even if the energy density is perfectly smooth the
reheating process will perturb all scales.

Nevertheless, if reheating were a purely classical process, then as
Zel'dovich (1965) and Zel'dovich and Novikov (1983) showed, the
rearrangement of energy could not yield a power spectral index less
than 4.  But reheating is not a classical process, and thus,
according to Gabrielli et al (2004), is not so constrained,
essentially because the correlations of quantum fields behave
differently from those of classical fields.  The real-space behavior
of $\delta_R$ described in eq.~(2) or (3) is mathematically
compatible with $P(k)\propto k^n$ for any $n$ in the range $1<n\le
4$.

Now in the synchronous gauge the perturbed metric in comoving
distance coordinates and conformal time is  \beq ds^2 = a^2 (\tau)
[-d\tau^2 + (\delta_{ij} + \h_{ij}) dx^i dx^j] \eeq ($d\tau =
dt/a(t)$ with $c=1$ for now).  The Planck function for the energy
density is, in this metric, the same as that in the unperturbed
metric where $\h_{ij} = 0$.  To see this, we start with the mean and
unperturbed occupation number $n({\bf p},T) = 1/[ \exp(p^0/k_{\rm
B}T) - 1]$ per quantum state of the primeval fireball, where ${\bf
p}$ is the physical momentum, and $p^0=|{\bf p}|$ the physical
energy.   When the metric is perturbed according to eq.~(4), we may
use the results of Ma \& Bertschinger (1995), who found that the
momenta conjugate to $x^i$ are $P_i = a(\delta_{ij} + \h_{ij}/2)p^j$
(and also $P_0 = -ap^0$), and that the mean of the momentum-energy
density tensor is
 \beq T_{\mu \nu} = 2\int
 (-g)^{-\frac{1}{2}} \frac{dP_1 dP_2 dP_3}{h^3} \frac{P_\mu
 P_\nu}{P^0} n({\bf P},T_0), \eeq
where $T_0 = aT$, $(-g)^{-\frac{1}{2}} = a^{-4} (1-\h/2)$ and $dP_1
dP_2 dP_3 = a^3 (1+\h/2) (p^0)^2 dp^0 d\Omega$, with $\h=\h_{ii}$,
where $d\Omega$ is the element of solid angle. It is evident that to
lowest order the effect of the metric fluctuation $\h$ cancels, so
that e.g.\ in the perturbed metric,
 \beq T^0{}_0 = -2\int\frac{(p^0)^3 dp^0 d\Omega}{h^3}
 \frac{1}{\exp\left(\frac{p^0}{k_{\rm B}T}\right) - 1}, \eeq
which is the same as that in the unperturbed background Friedmann
space-time.

From eq.~(6) ensues a `benchmark', or reference, level of spatial
fluctuations for the early radiation universe, viz.\ the spatial
thermal noise of an equilibrium photon gas of volume $V$ and radius
$R$ (both comoving), with standard deviation $\delta_R^{{\rm th}} =
(16k_{\rm B}/3S)^{1/2}$, where $S$ is the entropy in this volume.
Since the entropy is conserved during the later adiabatic expansion
of the universe, this is also the entropy in the same volume today,
namely $S=(43/22) \times (4u_0 V/T_0)$, where $u_0$ is the energy
density of a photon gas at $T_0 = 2.725$ K, and the factor of
$43/22$ arises because the entropy today includes not only that of
photons but also that of neutrinos. (The entropy of the photons
today is equal to that of photons plus electrons and positrons
before pair annihilation. Hence the ratio of the photon entropy to
neutrino entropy now is $(2+ 4 \times 7/8)/(6 \times 7/8)=22/21$.
See for example Peacock 1999, pp.~279--281.)  Thus we find
  \beq \delta_R^{{\rm th}}
  = \left(\frac{88 k_{\rm B}T_0}{43 u_0 V}\right)^{\frac{1}{2}} =
  \left(\frac{165 h^3 c^3}{43\pi^5 V k_{\rm B}^3 T_0^3}\right)^{\frac{1}{2}} =
  2.10 \times 10^{-2}
  \left(\frac{R}{1~{\rm cm}}\right)^{-\frac{3}{2}}, \eeq
Note that because this only depends on the entropy, which is
conserved, eq.~(7) does not involve the extra degrees of freedom
that may have arisen as a result of the proliferation of particle
species in the early universe when temperatures were very high. Thus
in the context of thermal noise the gas at any such epoch may
roughly be treated as an ensemble of randomly moving photons, with
Planckian energy spectrum.  This then affords the following
heuristic derivation of eq.~(7).  Observe that the mean photon
energy $\bar\epsilon=h\bar\nu$ and its standard deviation
$\delta\epsilon$ are both $\sim k_{\rm B} T_0$, so that for a volume
of total photon number $N$ and total energy $\bar E\sim Nk_{\rm B}
T_0$ the variance in $E$ is $(\delta E)^2 = N(\delta\epsilon)^2 \sim
N(k_{\rm B} T_0)^2$, or
 \beq \left(\frac{\delta E}{\bar E}\right)^2 =
 \left(\frac{\delta u}{u_0}\right)^2 \approx \frac{1}{N} =
 \frac{1}{8\pi V} \left(\frac{hc}{k_{\rm B} T_0}\right)^3
 \left(\int_0^\infty
 \frac{x^2 dx}{e^x -1}\right)^{-1}, \eeq
where the integral in parentheses equals 2.4. The square root of
this is only 15 \% larger than the correct answer of eq.~(7).

It is important to note that the part played in structure formation
by the perturbations of eq.~(7) has not been sufficiently
recognized. This is most easily seen by applying the linear growth
equation (eq.~5.123, Peebles 1993) to a small subhorizon region $R
\ll r_H$, for which expansion is negligible and the gravitational
term may be ignored.  In that case it reduces to the equation of
undamped sound wave propagation, predicting zero growth.  We know,
however, that on such small scales, where local thermodynamic
equilibrium (LTE) can quickly be restored by particle diffusion, any
super-thermal fluctuations are damped, and any sub-thermal ones are
enhanced, in relatively short times.

During the reheating phase between the cosmic times $t_i$ and $t_f$
a classical inflationary vacuum would have dissipated into heat.
Since the energy of this `false vacuum' is {\it uniform} at $t_i$,
the density contrast in superhorizon volumes will still be small at
$t_f$.  Nonetheless, the important point is that the contrast {\it
cannot be exactly zero} because thermal energy is not vacuum energy.
Once particles are present, then on the surface of each such volume
there exists a thin layer within which particles located outside
this volume can random walk into it, and vice versa.  The comoving
thickness of this layer, assuming $\delta t=t_f-t_i$ is not too
large, so that we may treat $H(t)$ as nearly constant and equal to
its radiation-era value at $t_f$, is
 \beq \delta R \approx \int_{t_i}^{t_f}\frac{cdt}{a(t)} \approx
 -\frac{c\delta z}{H(t_f)} \approx
 \frac{c(z_i-z_f)}{H_0 z_f^2} \sqrt{\frac{2}{g\Omega_\gamma}}, \eeq
where $g$ is the number of helicity states of light particles at the
time of reheating (fermions counting with a factor 7/8), and
$\Omega_\gamma$ is the normalized radiation density parameter
(including only photons, not neutrinos).  Thus the energy variance
is $(\delta E)^2 = N'(\delta\epsilon)^2 \approx N'(kT)^2$ where $N'
\approx 3N\delta R/R$.  The effect, as already explained, is
neglected by the linear perturbation theory.  Since the total energy
remains at $\bar E \sim NkT$ we now have
 \beq \delta_R = \frac{\delta u}{u_0} = \frac{\delta E}{\bar E}
 \approx \delta_R^{{\rm th}} \sqrt{\frac{N'}{N}}
 \approx \delta_R^{{\rm th}} \sqrt{\frac{3\delta R}{R}}
 \propto \frac{1}{R^2},~{\rm for}~R\geq r_H(t_i)\approx r_H(t_f), \eeq
where the last step was taken with the help of eq.~(7).

In this way, density seeds with the scaling relation of eq.~(2)
became available quite naturally. Specifically, $\delta_R^2 \propto
1/R^4$ corresponds mathematically to $n=1+\alpha$ where $\alpha$ is
arbitrarily positive.  Physically, as we remarked earlier, small
diffusive movements of particles \emph{in the quantum-interaction
regime} can yield a power spectrum with $P(k) \propto k^n$ and
$0<n\leq 4$. In the case of $n < 1$, eq. (2) says that $\delta_R^2
\propto 1/R^{3+n}$ for $R>r_H$, the scaling with $R$ now differs
from that of eq. (10).  This is not surprising, because eq. (10) was
derived assuming a classical horizon, which does not always lead to
the same $R$-dependence of $\delta_R$ as the scenario of the quantum
particle diffusion.  The recent \texttt{WMAP} data (Hinshaw et al
2009) give as the best fit $n$ the slightly lower value of $n=
0.96$, or $\delta_R^2 \propto 1/R^{3.96}$, very close to the form of
eq. (10).

Turning next to the question of amplitude, the factor that
determines the significance of the role of diffusion. As we have
emphasized, reheating is a quantum process, so ideally a full
quantum treatment would be required.  However, while a classical
calculation may not necessarily reveal the correlations inherent in
a quantum process, they should give a reasonable estimate of the
magnitude of the effect.  In reheating, the universe made a
transition from vacuum energy domination to radiation domination,
during which $r_H$ was changing slowly (from a decreasing function
of time to an increasing one). The value of $r_H$ for this period is
determined by the reheating temperature $T_{\rm reheat}$, which we
assume to be of order the GUT scale, i.e. one could adopt
 \beq r_H\approx \frac{c}{a_{{\rm reheat}} H_{\rm reheat}} =
 \frac{T_0}{T_{\rm reheat}}\frac{c}{H_0}
 \sqrt{\frac{2}{g\Omega_\gamma}} \approx
 200 \left(\frac{k_{{\rm B}} T_{\rm reheat}}{10^{15}~{\rm GeV}}\right)^{-1}
 ~{\rm cm}. \eeq
We may then use eqs.~(9) and (10) to obtain
 \beq \delta_R \approx 1.29 \times 10^{-5}
 \left(\frac{k_{\rm B}T_{\rm reheat}}
 {10^{15}~{\rm GeV}}\right)^{\frac{3}{2}}
 \left(\frac{g}{100}\right)^{\frac{3}{4}}
 \left(\frac{z_i - z_f}{z_f}\right)^{\frac{1}{2}}
 \left(\frac{R}{r_H}\right)^{-2}~{\rm for}~R\geq r_H. \eeq
It is reasonable to assume that $(z_i - z_f)/z_f \lesssim 1$ (which
just means reheating is not protracted), but that this ratio is not
\emph{very} small compared to unity.  In that case, we see that
$\delta_R$ for the horizon entry scale of $R=r_H$ {\it is} the
amplitude of $\approx 10^{-5}$ observed by \texttt{WMAP} and
required by structure formation models, i.e.\ eq.~(2).
By comparing eq.~(3) with eq.~(12), one sees that both the desired
shape and normalization of $\delta_R$ can emerge from this
mechanism,

For each superhorizon region enclosed by a real sphere the surface
diffusion will continue with time beyond $t=t_f$ as $\delta_R
\propto t^{1/4}$ during the radiation era, since $\delta_R^2 \propto
\delta R \propto t^{1/2}$ (eqs. (9) and (10) with $a(t) \propto
t^{1/2}$). This occurs independently of the gravitational growth of
linear theory, viz.\ $\delta_R \propto r_H^2 \propto a^2(t) \propto
t$ in the synchronous gauge, which commences at $t=t_f$ (when $r_H$
fully exhibits its radiation dominated time dependence) and persists
until re-entry.  Since the latter scales with time much faster than
the former, there is no need to take into account the $\delta_R
\propto t^{1/4}$ rate at epochs $t>t_f$.  The role played by the
diffusion process is really to `plant the seeds' during the period
$t_i < t < t_f$ when linear growth had yet to get under way, i.e.\
if initial density contrasts were absent by $t=t_f$, there would be
no seeds available for developing into the structures of today.

In conclusion, the mechanism we proposed will certainly generate
primordial density perturbations, and under very reasonable
assumptions these could have the required amplitude and spectral
index to agree with \texttt{WMAP} observations.  There is then no
need to postulate additional quantum fluctuations in the vacuum to
solve the `problem of seeds'.  As we have stressed, this is possible
only because reheating is an inherently quantum process.  At the
present time we do not have a specific quantum model to propose that
would have the desired properties.  Rather we assumed that such a
model exists, and pursued the consequences of this assumption.  If
the proposal stands up, then the development of a specific working
model will be an important task for the future.  If the ideas
presented here can actually account for all the essential parameters
by the time the data have the quality to clinch them, the search for
a simple and viable inflationary hypothesis might be freed from
constraints and greatly simplified.

The authors thank Ruth Durrer, Francesco Sylos Labini, and Robert
Crittenden for helpful discussions and the provision of vital
technical information.

\vspace{2mm}

\noindent {\bf References}

\noindent Albrecht, A.J. \& Steinhardt, P.J., 1982, PRL, 48, 1220.

\noindent Bennett, C.L., et al, 2003, ApJS, 148, 97.

\noindent Boyle, L.A., Steinhardt, P., \& Turok, N., 2006, PRL, 96,
111301.

\noindent Gabrielli, A., Joyce, M., \& Sylos Labini, F., 2002, PRD,
65, 083523.

\noindent  Gabrielli, A., Joyce, M., Marcos, B., \& Viot, P., 2004,
Europhys. Lett., 66, 1.

\noindent Guth, A.H., 1981, PRD, 23, 347.

\noindent Hall, L.J., \& Oliver, S., 2004, Nuclear Physics B
Proceedings Supplements, 137, 269.

\noindent Hinshaw, G. et al 2009, ApJS, 180, 225.

\noindent Kawasaki, M., Yamaguchi, M., \& Yokoyama, J., 2003, PRD,
68, 023508.

\noindent Kolb, E.W., \& Turner, M.S., 1990, The Early Universe,
Addison-Wesley.

\noindent Ma, C.-P., \& Bertschinger, E., 1995, ApJ, 455, 7.

\noindent Padmanabhan, T., 1993, Structure formation in the
universe, Cambridge University Press.

\noindent Peacock, J.A., 1999, Cosmological Physics, Cambridge
University Press.

\noindent Peebles, P.J.E., 1993, Principles of Physical Cosmology,
Princeton University Press.

\noindent Smoot G. et al., 1992, Ap.J., 396, L1.

\noindent Spergel, D.N. et al 2007, ApJS, 170, 377.

\noindent Vachaspati, T., \& Trodden, M., 2000, PRD, 61, 023502.

\noindent Yamaguchi, M., \& Yokoyama, J., 2003, PRD, 70, 023513.

\noindent Zel'dovich, Ya., 1965, Adv. Astron. Ap. 3, 241.

\noindent Zel'dovich, Ya., \& Novikov, I., 1983, Relativistic
Astrophysics, Vol. 2, Univ. Chicago Press.

\end{document}